\documentclass[12pt]{article}

\usepackage{a4wide}
\usepackage[pdftex,usenames,dvipsnames]{color}
\usepackage{graphics}
\usepackage{amsfonts}
\usepackage{amssymb}
\usepackage{accents}

\newcommand{\nit}{\noindent}

\newcommand{\np}{\newpage}
\newcommand{\dsp}{\displaystyle}
\newcommand{\vs}[1]{\vspace{#1 ex}}
\newcommand{\hs}[1]{\hspace{#1 em}}
\newcommand{\bfr}{\begin{flushright}}
\newcommand{\efr}{\end{flushright}}
\newcommand{\bc}{\begin{center}}
\newcommand{\ec}{\end{center}}
\newcommand{\ben}{\begin{enumerate}}
\newcommand{\een}{\end{enumerate}}

\newcommand{\be}{\begin{equation}}
\newcommand{\ee}{\end{equation}}
\newcommand{\ba}{\begin{array}}
\newcommand{\ea}{\end{array}}
\newcommand{\ct}{\cite}
\newcommand{\bit}{\bibitem}

\newcommand{\del}{\delta}

\newcommand{\ve}{\varepsilon}

\newcommand{\kg}{\kappa}
\newcommand{\lb}{\lambda}
\newcommand{\sg}{\sigma}
\newcommand{\rg}{\rho}

\newcommand{\Gam}{\Gamma}

\newcommand{\bfer}{\bold{r}}

\newcommand{\bfx}{\bold{x}}

\newcommand{\bfB}{\bold {B}}

\newcommand{\bfE}{\bold {E}}
\newcommand{\bfF}{\bold {F}}

\newcommand{\bfP}{\bold {P}}

\newcommand{\hh}{\hat{h}}

\newcommand{\hr}{\hat{r}}

\newcommand{\hT}{\hat{T}}

\newcommand{\lh}{\left(}
\newcommand{\rh}{\right)}
\newcommand{\ld}{\left.}
\newcommand{\rd}{\right.}

\newcommand{\nb}{\nabla}

\newcommand{\der}{\partial}

\newcommand{\Tr}{\mbox{Tr}}

\begin{document}

\pagestyle{empty}

\bfr
Nikhef 2022-019
\efr

\vs{7}
\bc
{\bf \large Curvature dynamics in General Relativity}
\vs{7} 

{\large J.W.\ van Holten}\\
\vs{3}

{\large Nikhef, Amsterdam NL}\\
\vs{2}
and \\
\vs{2} 
{\large Leiden University, Leiden NL}
\vs{3} 

Nov.\ 8, 2022
\ec
\vs{5}

\nit
{\small 
{\bf Abstract} \\
The equations of General Relativity are recast in the form of a wave equation for the Weyl tensor. 
This allows to reformulate gravitational wave theory in terms of curvature waves, rather than 
metric waves. The existence of two transverse polarization states for curvature waves is proven 
and in the linearized approximation the quadrupole formula is rederived. A perturbative scheme  
to extend the linearized result to the non-linear regime is outlined. }
\np
~ 

\np

\pagestyle{plain}
\pagenumbering{arabic}

\nit
{\bf \large 1.\ Introduction} 
\vs{1}

\nit
The essential content of General Relativity (GR) resides in the identification of gravity with the
geometric property of space-time curvature \ct{einstein:1916}. All observer frames in free fall 
can be identified with local inertial frames in which gravity is absent, as the local space-time 
geometry is flat. But the relative acceleration between local inertial frames at different points 
in space at different times is encoded in the space-time curvature and cannot be eliminated 
by any choice of reference frame. 

Therefore an essential description of gravitation is to be cast in terms of the dynamics of 
curvature. It is the aim of this paper to provide such a description and to show how some 
familiar results of GR describing observed gravitational phenomena can be rederived in 
such a frame work. 
\vs{2}

\nit
{\bf \large 2.\ Space-time curvature} 
\vs{1}

\nit
In this section we summarize the properties of space-time curvature and establish our 
notation.

In geometry curvature is measured by the extent to which parallel displacements of vectors
and higher-rank tensors in two independent directions commute. In differential form this is 
expressed by the Ricci identity, which when implemented on a covariant vector field $V_{\mu}(x)$
takes the form
\be
\left[ \nb_{\mu}, \nb_{\nu} \right] V_{\kg} = - R_{\mu\nu\kg}^{\;\;\;\;\;\lb}\, V_{\lb},
\label{2.1}
\ee
where the coefficients $R_{\mu\nu\lb}^{\;\;\;\;\;\kg}$ are the components of the Riemann 
curvature tensor. This identity relates the Riemann curvature tensor to the covariant derivative 
$\nb_{\mu}$ and the associated connection with components $\Gam_{\mu\nu}^{\;\;\;\lb}$:
\be
R_{\mu\nu\kg}^{\;\;\;\;\;\lb} = ( \der_{\mu} \Gam_{\nu} - \der_{\nu} \Gam_{\mu} - 
 \left[ \Gam_{\mu}, \Gam_{\nu} \right] )_{\kg}^{\;\,\lb}.
\label{2.2}
\ee
The standard choice of connection is the one which transports the metric parallel to itself: 
\be
\nb_{\lb} g_{\mu\nu} = 0.
\label{2.3}
\ee
This condition is sometimes known as the metric postulate; it results in the Riemann-Chistoffel 
connection 
\be
\Gam_{\mu\nu}^{\;\;\;\lb} = \frac{1}{2}\, g^{\lb\kg} \lh \der_{\mu} g_{\kg\nu} + \der_{\nu} g_{\mu\kg} 
 - \der_{\kg} g_{\mu\nu} \rh.
\label{2.4}
\ee
By the definition (\ref{2.1}) and applying the Ricci identity to the metric postulate, one 
establishes the symmetry properties of the Riemann tensor in the fully covariant representation
\be
R_{\mu\nu\kg\lb} = - R_{\nu\mu\kg\lb} = - R_{\mu\nu\lb\kg} = R_{\kg\lb\mu\nu}, 
\label{2.5}
\ee
and the cyclic property 
\be
R_{\mu\nu\kg\lb} + R_{\nu\kg\mu\lb} + R_{\kg\mu\nu\lb} = 0. 
\label{2.6}
\ee
In addition to these algebraic indentities the cyclic Jacobi-identity for three covariant derivatives
guarantees the Bianchi identity for the Riemann tensor:
\be
\nb_{\sg} R_{\mu\nu\kg\lb} + \nb_{\mu} R_{\nu\sg\kg\lb} + 
 \nb_{\nu} R_{\sg\mu\kg\lb} = 0.
\label{2.7}
\ee
By contraction with the inverse metric $g^{\sg\lb}$ this identity implies another one for the 
divergence of the Riemann tensor: 
\be
\nb^{\lb}R_{\lb\kg\mu\nu} = \nb_{\mu} R_{\nu\kg} - \nb_{\nu} R_{\mu\kg}.
\label{2.7a}
\ee
In view of its symmetry properties the Riemann tensor in 4 space-time dimensions has 20 
independent components. Of these 10 are contained in the trace of the Riemann tensor,
the symmetric Ricci tensor 
\be
R_{\mu\nu} = R_{\nu\mu} = R_{\mu\lb\nu}^{\;\;\;\;\;\lb}.
\label{2.8}
\ee
The trace of the Ricci tensor is the Riemann curvature scalar $R = R_{\mu}^{\;\,\mu}$. 
The other 10 components of the Riemann tensor are contained in its traceless part, known 
as the Weyl tensor \ct{weyl:1918}, with components
\be
W_{\mu\nu\kg\lb} = R_{\mu\nu\kg\lb} - \frac{1}{2} \lh g_{\mu\kg} R_{\nu\lb} 
 - g_{\mu\lb} R_{\nu\kg} - g_{\nu\kg} R_{\mu\lb} + g_{\nu\lb} R_{\mu\kg} \rh + 
 \frac{1}{6} \lh g_{\mu\kg} g_{\nu\lb} - g_{\mu\lb} g_{\nu\kg} \rh R. 
\label{2.9}
\ee
Its is straightforward to check that the Weyl tensor has the same algebraic symmetry properties 
(\ref{2.5}, \ref{2.6}) as the Riemann tensor, and that in addition its trace vanishes: 
\be
W_{\mu\nu\kg}^{\;\;\;\;\;\nu} = 0.
\label{2.10}
\ee
Indeed, this condition eliminates 10 of the original 20 components of the Riemann tensor, leaving
10 other components as claimed. 

Taking traces of the Bianchi identity (\ref{2.7}) and its divergence (\ref{2.7a}), it follows that the 
Ricci tensor has a divergence-free extension, the Einstein tensor: 
\be
G_{\mu\nu} = R_{\mu\nu} - \frac{1}{2}\, g_{\mu\nu} R, \hs{2} \nb^{\mu} G_{\mu\nu} = 0.
\label{2.11}
\ee
The Riemann tensor can now be decomposed into the Weyl tensor, the Einstein tensor and 
the Riemann scalar, by inverting (\ref{2.9}): 
\be
R_{\mu\nu\kg\lb} = W_{\mu\nu\kg\lb} + \frac{1}{2} \lh g_{\mu\kg} G_{\nu\lb} - g_{\mu\lb} G_{\nu\kg} 
 - g_{\nu\kg} G_{\mu\lb} + g_{\nu\lb} G_{\mu\kg} \rh + \frac{1}{3} \lh g_{\mu\kg} g_{\nu\lb} - 
  g_{\mu\lb} g_{\nu\kg} \rh R.
\label{2.12}
\ee
The part of curvature dynamics determined by the physical content of the universe formed 
by matter and radiation is described by the Einstein equations
\be
G_{\mu\nu} = - 8 \pi G T_{\mu\nu},
\label{2.13}
\ee
with $T_{\mu\nu}$ the energy-momentum tensor of matter and radiation and $G$ the gravitational 
constant; note that we use units in which the speed of light $c = 1$. The condition (\ref{2.11})
of vanishing divergence of the Einstein tensor thereby becomes a consistency requirement for 
the local conservation of energy and momentum. 

Taking the trace of the Einstein tensor it follows that 
\be
G_{\mu}^{\;\mu} = - R = - 8 \pi G T,
\label{2.14}
\ee
using the notation $T = T_{\mu}^{\;\mu}$ for the trace. The general expression for the Riemann 
curvature in regions with energy-momentum density $T_{\mu\nu}$ therefore is 
\be 
R_{\mu\nu\kg\lb} = W_{\mu\nu\kg\lb}  - 8 \pi G \lh \frac{1}{2} \lh g_{\mu\kg} T_{\nu\lb} 
 - g_{\mu\lb} T_{\nu\kg} - g_{\nu\kg} T_{\mu\lb} + g_{\nu\lb} T_{\mu\kg} \rh - 
 \frac{1}{3} \lh g_{\mu\kg} g_{\nu\lb} - g_{\mu\lb} g_{\nu\kg} \rh T \rh,
\label{2.15}
\ee
which in the vacuum (empty space-time regions where $T_{\mu\nu} = 0$) reduces to the 
Weyl tensor. The Bianchi identity (\ref{2.7}) can be rewritten as an identity for the Weyl 
tensor; it then takes the form
\be
\nb_{\left[ \sg \rd} W_{\ld \mu\nu \right] \kg \lb} =    
 - g_{\kg\left[\mu \rd} \nb_{\sg} G_{\ld \nu \right]\lb} + 
 g_{\lb\left[\mu \rd} \nb_{\sg} G_{\ld \nu \right]\kg} 
 - \frac{2}{3}\, g_{\kg \left[ \mu \rd} \nb_{\sg} R\, g_{\ld \nu \right] \lb},
\label{2.16}
\ee
where the square brackets denote complete anti-symmetrization of the enclosed indices
$[\mu\sg\nu]$ with total weight equal to one. By the same substitution (\ref{2.13}, \ref{2.14}) 
it follows that the right-hand side vanishes in the vacuum. Finally, by contraction with 
$g^{\sg\lb}$ we find a result analogous to (\ref{2.7a}) for the divergence of the Weyl tensor
\ct{schouten:1951,carroll:2004}:
\be
\nb^{\lb} W_{\lb\kg\mu\nu} = \frac{1}{2} \lh \nb_{\mu} G_{\nu\kg} - \nb_{\nu} G_{\mu\kg} \rh
 + \frac{1}{6} \lh g_{\kg\nu} \nb_{\mu} R - g_{\kg\mu} \nb_{\nu} R \rh.
\label{2.17}
\ee
\vs{2}

\nit
{\bf \large 3.\ Curvature dynamics}  
\vs{1}

\nit
In the previous section we have established results for the mathematical and physical 
properties of space-time curvature, with the Riemann tensor expressing the overall 
curvature in the presence of energy densities due to matter and radiation, whilst the 
Weyl tensor expresses the purely gravitational contribution to the curvature as it exists 
in a vacuum. In this section we discuss the dynamics of these curvature tensors 
themselves, as follow from these properties. 

In the case of gravity it is simpler to derive the equation for the overall curvature 
in the presence of energy densities (the Riemann curvature) then for pure gravity
(the Weyl curvature). The derivation starts from the Bianchi identity (\ref{2.7}),
taking a divergence:
\be
\nb^2 R_{\mu\nu\kg\lb} - \nb^{\sg} \nb_{\mu} R_{\sg\nu\kg\lb} + 
 \nb^{\sg} \nb_{\nu} R_{\sg\mu\kg\lb} = 0,
\label{3.1}
\ee
and using the Ricci identity and the result (\ref{2.7a}) for the divergence of the Riemann
tensor, to end up with 
\be
\ba{l}
\nb^2 R_{\mu\nu\kg\lb} - 2 R_{\mu\sg\kg}^{\;\;\;\;\;\rg} R_{\lb\rg\nu}^{\;\;\;\;\;\sg} 
 + 2 R_{\mu\sg\lb}^{\;\;\;\;\;\rg} R_{\kg\rg\nu}^{\;\;\;\;\;\sg} + 
 R_{\mu\nu\rg}^{\;\;\;\;\;\sg} R_{\kg\lb\sg}^{\;\;\;\;\;\rg}  = \\
 \\
\dsp{\hs{2} =\, - \frac{1}{2} \lh R_{\mu\nu\kg\rg} R^{\rg}_{\;\,\lb} - R_{\mu\nu\lb\rg} R^{\rg}_{\;\,\kg}
  + R_{\kg\lb\mu\rg} R^{\rg}_{\;\,\nu} - R_{\kg\lb\nu\rg} R^{\rg}_{\;\,\mu} \rh }\\
 \\
\dsp{ \hs{3.4} +\, \frac{1}{2} \lh \left\{ \nb_{\mu}, \nb_{\kg} \right\} R_{\nu\lb} - 
  \left\{ \nb_{\mu}, \nb_{\lb} \right\} R_{\nu\kg} - \left\{ \nb_{\nu}, \nb_{\kg} \right\} R_{\mu\lb} + 
  \left\{ \nb_{\nu}, \nb_{\lb} \right\} R_{\mu\kg} \rh, }
\ea
\label{3.2}
\ee
where the curly braces denote the symmetric anti-commutator of the covariant derivatives 
enclosed. Note that the right-hand side vanishes in a vacuum environment, in which case the 
Riemann tensor can be replaced by the Weyl tensor. Indeed, written in terms of the  Weyl 
tensor the full equation becomes 
\be
\ba{l}
\nb^2 W_{\mu\nu\kg\lb} - 2 W_{\mu\sg\kg}^{\;\;\;\;\;\rg} W_{\lb\rg\nu}^{\;\;\;\;\;\sg} 
 + 2 W_{\mu\sg\lb}^{\;\;\;\;\;\rg} W_{\kg\rg\nu}^{\;\;\;\;\;\sg} - 
 W_{\mu\nu\rg}^{\;\;\;\;\;\sg} W_{\kg\lb\sg}^{\;\;\;\;\;\rg}  = \\
 \\
\dsp{ \hs{1} =\, - \frac{1}{2} \lh W_{\mu\nu\kg\rg} G^{\rg}_{\;\,\lb} - W_{\mu\nu\lb\rg} G^{\rg}_{\;\,\kg}
 + W_{\kg\lb\mu\rg} G^{\rg}_{\;\,\nu} - W_{\kg\lb\nu\rg} G^{\rg}_{\;\,\mu} \rh - R W_{\mu\nu\kg\lb} }\\
 \\
\dsp{ \hs{2.4} + \lh g_{\mu\kg} W_{\nu\rg\lb\sg} - g_{\mu\lb} W_{\nu\rg\kg\sg} 
 - g_{\nu\kg} W_{\mu\rg\lb\sg} + g_{\nu\lb} W_{\mu\rg\kg\sg} \rh G^{\rg\sg} }\\
 \\
\dsp{ \hs{2.4} - \lh g_{\mu\kg} [G^2]_{\nu\lb} - g_{\mu\lb} [G^2]_{\nu\kg}
 - g_{\nu\kg} [G^2]_{\mu\lb} + g_{\nu\lb} [G^2]_{\mu\kg} \rh 
 + \frac{1}{2} \lh g_{\mu\kg} g_{\nu\lb} - g_{\mu\lb} g_{\nu\kg} \rh \Tr\, G^2 }\\
 \\
\dsp{ \hs{2.4} - \frac{2}{3}\, R \lh g_{\mu\kg} G_{\nu\lb} - g_{\mu\lb} G_{\nu\kg} - 
 g_{\nu\kg} G_{\mu\lb} + g_{\nu\lb} G_{\mu\kg} \rh - 
 \frac{1}{3}\, R^2 \lh g_{\mu\kg} g_{\nu\lb} - g_{\mu\lb} g_{\nu\kg} \rh }\\
 \\
\dsp{ \hs{2.4} +\, \frac{1}{2} \lh \left\{ \nb_{\mu}, \nb_{\kg} \right\} G_{\nu\lb} - 
  \left\{ \nb_{\mu}, \nb_{\lb} \right\} G_{\nu\kg} - \left\{ \nb_{\nu}, \nb_{\kg} \right\} G_{\mu\lb} 
  + \left\{ \nb_{\nu}, \nb_{\lb} \right\} G_{\mu\kg} \rh }\\
 \\
\dsp{ \hs{2.4} -\, \frac{1}{2} \lh g_{\mu\kg} \nb^2 G_{\nu\lb} - g_{\mu\lb} \nb^2 G_{\nu\kg} 
 - g_{\nu\kg} \nb^2 G_{\mu\lb} + g_{\nu\lb} \nb^2 G_{\mu\kg} \rh
}\\
 \\
\dsp{ \hs{2.4} +\, \frac{1}{2} \lh g_{\nu\mu} \nb_{\kg} \nb_{\lb} R - g_{\nu\lb} \nb_{\kg} \nb_{\mu} R
 - g_{\kg\mu} \nb_{\nu} \nb_{\lb} R + g_{\kg\lb} \nb_{\nu} \nb_{\mu} R \rh }\\
 \\
\dsp{ \hs{2.4} -\, \frac{1}{3} \lh g_{\mu\kg} g_{\nu\lb} - g_{\mu\lb} g_{\nu\kg} \rh \nb^2 R. }\\
\ea 
\label{3.3}
\ee
As the wave equations for the curvature tensors are derived from the Ricci and Bianchi identities,
their general solution is given by equations (\ref{2.2}) and (\ref{2.4}). Specific expressions for the 
metric and connection are implied by imposing the Einstein equations (\ref{2.13}) as a constraint 
on $G_{\mu\nu}$. Therefore equations (\ref{3.2}), (\ref{3.3}) are automatically satisfied by all 
solutions of the Einstein equations.

Eq.\ (\ref{3.3}) is a non-linear wave equation for the Weyl curvature tensor. Given a particular 
vacuum metric --making the right-hand side of this equation to vanish-- it defines an extremum, 
under free variations of the tensor $W_{\mu\nu\kg\lb}$, of the action functional 
\be
\ba{lll}
S[W;g] & = & \dsp{ \int d^4x \sqrt{-g} \left[ - \frac{1}{2}\, \nb^{\rg} W^{\mu\nu\kg\lb} 
 \nb_{\rg} W_{\mu\nu\kg\lb} - \frac{1}{3}\, W_{\mu\nu}^{\;\;\,\kg\lb} W_{\kg\lb}^{\;\;\,\rg\sg} 
 W_{\rg\sg}^{\;\;\,\mu\nu} \rd }\\
 & & \\
 & & \dsp{ \ld -\, \frac{4}{3}\, W^{\mu\;\,\kg}_{\;\,\nu\;\,\lb} W_{\mu\;\,\kg}^{\;\,\sg\;\,\rg} 
  W^{\;\,\nu\;\,\lb}_{\sg\;\,\rg} \right]. }
\ea
\label{3.4}
\ee
A brief discussion and generalization of this action is presented in appendix A. 
\vs{2}

\np
\nit
{\bf \large 4.\ Curvature polarization modes} 
\vs{1}

\nit
In a vacuum the dynamical solutions of the curvature equations 
\be
\nb^2 W_{\mu\nu\kg\lb} - 2 W_{\mu\sg\kg}^{\;\;\;\;\;\rg} W_{\lb\rg\nu}^{\;\;\;\;\;\sg} 
 + 2 W_{\mu\sg\lb}^{\;\;\;\;\;\rg} W_{\kg\rg\nu}^{\;\;\;\;\;\sg} - 
 W_{\mu\nu\rg}^{\;\;\;\;\;\sg} W_{\kg\lb\sg}^{\;\;\;\;\;\rg}  = 0,
\label{4.1}
\ee
describe gravitational waves as waves of curvature. These equations follow from the Bianchi and 
Ricci identities, in particular (\ref{2.16}) and (\ref{2.17}), which in vacuum reduce to  
\be
\nb_{\sg} W_{\mu\nu\kg\lb} + \nb_{\mu} W_{\nu\sg\kg\lb} + \nb_{\nu} W_{\sg\mu\kg\lb} = 0, 
 \hs{2} \nb^{\lb} W_{\lb\kg\mu\nu} = 0.
\label{4.2}
\ee
Recalling also the cyclic property of the curvature tensor 
\[
W_{\mu\nu\kg\lb} + W_{\nu\kg\mu\lb} + W_{\kg\mu\nu\lb} = 0,
\]
we can now establish that curvature waves have precisely 2 polarization modes. 
To show this it is convenient to make a 3+1 space-time split, and define
\be
E_{ij} = W_{0i0j}, \hs{2} B_i^{\;j} = \frac{1}{2\sqrt{-g}}\, \ve^{0jmn} W_{0imn}, \hs{2}
 P^{ij} = - \frac{1}{4g}\, \ve^{0imn} \ve^{0jkl} W_{mnkl},
\label{4.3}
\ee
where the latin indices $k,l,m,n = (1,2,3)$ denote components of spatial 3-tensors.
All three spatial tensors are traceless: 
\be
\ba{l}
\dsp{ E_j^{\;j} = g^{ij} W_{0i0j} = g^{\mu\nu} W_{0\mu 0\nu} = 0, \hs{1} 
B_j^{\;j} = \frac{1}{2\sqrt{-g}}\, \ve^{0\mu\nu\lb} W_{0\mu\nu\lb} = 0, }\\
 \\
\dsp{ P_j^{\;j} = - \frac{1}{4g}\, g_{\rg\sg} \ve^{0\rg\mu\nu} \ve^{0\sg\kg\lb} W_{\mu\nu\kg\lb} = 0, }
\ea
\label{4.4}
\ee
as a result of the 4-dimensional tracelessness and cyclic property of the Weyl tensor. 
Next, in empty space-time the 3-divergence of these tensors vanishes: 
\be
\ba{l}
\dsp{ \nb^i E_{ij} = \nb^{\mu} W_{0\mu0j} = 0, \hs{1} 
\nb^i B_i^{\;j} = \frac{1}{2\sqrt{-g}}\, \ve^{0j\kg\lb} \nb^{\mu} W_{0\mu\kg\lb} = 0, }\\
 \\
\dsp{ \nb_i P^{ij} = - \frac{1}{4g}\, \ve^{0j\mu\nu} \ve^{0\sg\kg\lb} \nb_{\sg} W_{\kg\lb\mu\nu} = 0. }
\ea
\label{4.5}
\ee
Therefore as 3-tensors $E_{ij}$ and $P^{ij}$ are symmetric, traceless and divergence-free,
implying that they have only 2 independent components. Finally, these 3-tensors are related 
in terms of time-derivatives
\be
\nb_0 P^{ij} = - \frac{1}{4g}\, \ve^{0imn} \ve^{0jkl} \nb_0 W_{mnkl}
 = - \frac{1}{2g}\, \ve^{0imn} \ve^{0jkl} \nb_m W_{n0kl} 
 = \frac{1}{\sqrt{-g}}\, \ve^{0imn} \nb_m B_n^{\;j}, 
\label{4.6}
\ee
and
\be
\nb^0 B_i^{\;j} = \frac{1}{2\sqrt{-g}}\, \ve^{0jmn} \nb^0 W_{0imn} = 
 - \frac{1}{2\sqrt{-g}}\, \ve^{0jmn} \nb^k W_{kimn} = \sqrt{-g}\, \ve_{0ikl} \nb^k P^{lj}.
\label{4.7}
\ee
Similarly
\be
\nb_0 B_i^{\;j}  = \frac{1}{2\sqrt{-g}}\, \ve^{0j\mu\nu} \nb_0 W_{\mu\nu 0i} = 
 \frac{1}{\sqrt{-g}}\, \ve^{0j\mu\nu} \nb_{\nu} W_{0\mu 0i} = \frac{1}{\sqrt{-g}}\, \ve^{0jkl} \nb_{l} E_{ki},
\label{4.8}
\ee
and finally
\be
\nb^0 E_{ij} = \nb^0 W_{0i0j} = - \nb^k W_{ki0j} = \frac{1}{2}\, \ve_{0lki} \ve^{0lmn} \nb^k W_{0jmn} 
 = \sqrt{-g}\, \ve_{0lki} \nb^k B_j^{\;l}. 
\label{4.9}
\ee
Thus $B_i^{\;j}$ and $E_{ij}$ describe magnetic and electric components of the curvature, 
encoding the time-evolution of the two physical degrees of freedom contained in $P^{ij}$. 
These two degrees of freedom in $P^{ij}$ represent the independent physical spatial 
polarization components of the Weyl tensor.  
\vs{2}

\nit
{\bf \large 5.\ Weak gravity: the linear approximation}
\vs{1}

\nit
The curvature dynamics in general relativity simplifies considerably in the weak gravity 
limit, in which metric and curvature fluctuations can be treated in the linearized approximation
on a flat background space-time. The starting point of this linear theory is to split the metric 
$g_{\mu\nu}$ into a constant flat Minkowski background plus metric fluctuations: 
\be
g_{\mu\nu}(x) = \eta_{\mu\nu} + 2 h_{\mu\nu}(x). 
\label{5.1}
\ee
It is possible that such a single split can be made only in a restricted part of space-time,
because the local co-ordinates $x^{\mu}$ can not be extended to all of space-time.
If in a neighboring part of space-time, overlapping only partly with the previous one, a
different local set of co-ordinates $x^{\prime\,\mu}$ is necessary, such that in the domain of 
overlap of the co-ordinate systems
\be
x^{\prime\,\mu} - x^{\mu} = - 2 \xi^{\mu}(x),
\label{5.2}
\ee
then to first order in $\xi$ the split of the corresponding new metric $g'_{\mu\nu}(x')$ in the 
domain of overlap is 
\be
g'_{\mu\nu}(x) = \eta_{\mu\nu} + 2 h'_{\mu\nu}(x), \hs{2} 
 h'_{\mu\nu} = h_{\mu\nu} + \nb_{\mu} \xi_{\nu} + \nb_{\nu} \xi_{\mu}.
\label{5.3}
\ee
The linear approximation restricts all expressions for geometric quantities to the part linear 
in the fluctuations $h_{\mu\nu}$; thus for the inverse metric
\be
g^{\mu\nu} = \eta^{\mu\nu} - 2 h^{\mu\nu} + {\cal O}[h^2], \hs{2} 
h^{\mu\nu} \equiv \eta^{\mu\kg} h_{\kg\lb} \eta^{\lb\nu}, 
\label{5.4}
\ee
for the connection
\be
\Gam_{\mu\nu}^{\;\;\;\lb} = \eta^{\lb\kg} \lh \der_{\mu} h_{\nu\kg} + \der_{\nu} h_{\mu\kg} 
 - \der_{\kg} h_{\mu\nu} \rh + {\cal O}[h^2],
\label{5.5}
\ee
and for Riemann curvature
\be
R_{\mu\nu\kg\lb} = \der_{\nu} \der_{\lb} h_{\mu\kg} - \der_{\nu} \der_{\kg} h_{\mu\lb} 
 - \der_{\mu} \der_{\lb} h_{\nu\kg} + \der_{\mu} \der_{\kg} h_{\nu\lb} + {\cal O}[h^2]. 
\label{5.6}
\ee
Note that in this approximation the co- and contravariant components of vectors and tensors 
are always related by contraction with the Minkowski metric $\eta_{\mu\nu}$ and its inverse 
$\eta^{\mu\nu}$. Furthermore the co-ordinate transformations (\ref{5.2}) now reduce to 
gauge transformations 
\be
h'_{\mu\nu} = h_{\mu\nu} + \der_{\mu} \xi_{\nu} + \der_{\nu} \xi_{\mu}, \hs{2} 
\Gam_{\mu\nu}^{\prime\;\;\lb} = \Gam_{\mu\nu}^{\;\;\;\lb} + 2 \der_{\mu} \der_{\nu} \xi^{\lb},
\label{5.7}
\ee
which leave the Riemann tensor invariant: $R'_{\mu\nu\kg\lb} = R_{\mu\nu\kg\lb}$. 

The expressions for the Ricci tensor and Riemann scalar reduce in the linear approximation to
\be
\ba{l}
R_{\mu\nu} = \Box h_{\mu\nu} - \der_{\mu} \der^{\lb} h_{\lb\nu} - \der_{\nu} \der^{\lb} h_{\lb\mu}
 + \der_{\mu} \der_{\nu} h_{\lb}^{\;\lb}, \\
 \\
R = 2 \lh \Box h_{\mu}^{\;\mu} - \der^{\mu} \der^{\nu} h_{\mu\nu} \rh,
\ea
\label{5.8}
\ee
where $\Box$ is the d'Alembertian in Minkowski space. Therefore the Einstein 
equation becomes 
\be
G_{\mu\nu} = \Box h_{\mu\nu} - \der_{\mu} \der^{\lb} h_{\lb\nu} - \der_{\nu} \der^{\lb} h_{\lb\mu}
 + \der_{\mu} \der^{\nu} h_{\lb}^{\;\lb} - \eta_{\mu\nu} \lh \Box h_{\lb}^{\;\lb} - \der^{\kg} \der^{\lb} h_{\kg\lb} \rh
  = - 8 \pi G T_{\mu\nu},
\label{5.9}
\ee
where the linearized Einstein tensor satisfies the conservation law 
\be
\der^{\mu} G_{\mu\nu} = - 8 \pi G\, \der^{\mu} T_{\mu\nu} = 0.
\label{5.10}
\ee
The differential identities (\ref{2.7}), (\ref{2.7a}) satisfied by the Riemann tensor simplify to 
\be
\der_{\sg} R_{\mu\nu\kg\lb} + \der_{\mu} R_{\nu\sg\kg\lb} + \der_{\nu} R_{\sg\mu\kg\lb} = 0, 
 \hs{2} \der^{\mu} R_{\mu\nu\kg\lb} = - 8 \pi G \lh \der_{\kg} \hT_{\nu\lb} - \der_{\lb} T_{\nu\kg} \rh,
\label{5.11}
\ee
where 
\be
\hT_{\mu\nu} = T_{\mu\nu} - \frac{1}{2}\, \eta_{\mu\nu} T.
\label{5.13}
\ee
These equations now imply the inhomogeneous linear wave equation\footnote{This linear form of
the equation was derived before in ref.\ \ct{bieri:2014}}
\be
\Box R_{\mu\nu\kg\lb} = - 8 \pi G \lh \der_{\mu} \der_{\kg} \hT_{\nu\lb} - \der_{\mu} \der_{\lb} \hT_{\nu\kg}
 - \der_{\nu} \der_{\kg} \hT_{\mu\lb} + \der_{\nu} \der_{\kg} \hT_{\mu\lb} \rh,
\label{5.12}
\ee
Following the decomposition of the Weyl tensor in section 4 we introduce a $3 + 1$ decomposition 
of the linearized Riemann tensor: 
\be
E_{ij} = R_{0i0j}, \hs{1} B_{ij} = - \frac{1}{2}\, \ve_{jmn} R_{0imn}, \hs{1}
P_{ij} = \frac{1}{4}\, \ve_{ikl} \ve_{jmn} R_{klmn},
\label{5.14}
\ee
By the Bianchi identity, the cyclic property of the Riemann tensor and the Einstein equations it
follows that 
\be
\ba{l}
E_{jj} = - 8 \pi G\, \hT_{00} = - 4 \pi G \lh T_{jj} + T_{00} \rh, \hs{1} B_{jj} = 0, \\
 \\
P_{jj} = - 4 \pi G \lh \hT_{jj} + \hT_{00} \rh = - 8\pi G\, T_{00}.
\ea
\label{5.15}
\ee
Also
\be
\ba{ll}
\der_i E_{ij} = - 8\pi G \lh \der_j \hT_{00} - \der_0 \hT_{0j} \rh, & \der_j B_{ij} = 0,  \\
 \\
\der_i B_{ij} = - 8 \pi G\, \ve_{jkl} \der_k \hT_{l0}, & \der_i P_{ij} = 0,
\ea
\label{5.16}
\ee
and
\be
\ba{ll}
\ve_{ikl} \der_k E_{lj} = - \der_t B_{ij}, & 
\ve_{ikl} \der_k B_{lj} = \der_t E_{ij} + 8\pi G \lh \der_j \hT_{i0} - \der_0 \hT_{ij} \rh, \\
 \\
\ve_{jkl} \der_k B_{il} = - \der_t P_{ij}, &
\ve_{ikl} \der_k P_{lj} = \der_t B_{ij} + 8\pi G\, \ve_{jkl} \der_k \hT_{li}.
\ea
\label{5.17}
\ee
Thus in a vacuum all 3-tensor fields $\bfF = (\bfE, \bfB, \bfP)$ are traceless and transverse, and they satisfy the
homogeneous wave equation $\Box \bfF = 0$.
\vs{2}

\nit
{\bf \large 6.\ Static curvature geometries}
\vs{1}

\nit
As is well-known, the gauge transformations (\ref{5.7}) can be used to fix 
$h_{\mu\nu}(x_0) = \Gam_{\lb\nu}^{\;\;\;\mu}(x_0) = 0$ at a given point with co-ordinates 
$x^{\mu}_0$; this is achieved by taking 
\be
\xi_{\mu} =  \frac{1}{2}\, h_{\mu\nu}(x_0) \lh x^{\nu} - x_0^{\nu} \rh - 
 \frac{1}{4} \Gam_{\lb\nu\mu}(x_0) \lh x^{\lb} - x_0^{\lb} \rh \lh x^{\nu} - x_0^{\nu} \rh + 
 {\cal O}[(x - x_0)^3].
\label{6.1}
\ee
In this way a locally flat co-ordinate system in the neighborhood of $x_0$ is contructed. 
Of course, the curvature as encoded by the Riemann tensor can not be transformed away, 
as the Riemann tensor in gauge invariant. In a locally flat co-ordinate system with $x_0$ as 
the origin, such a geometry with {\em constant} Riemann tensor in the neighborhood of the 
origin is described by the tensor field 
\be
h_{\mu\nu}(x) = \frac{1}{6}\, R_{\mu\kg\nu\lb} x^{\kg} x^{\lb}, \hs{2} \der_{\sg} R_{\mu\kg\nu\lb} = 0.
\label{6.2}
\ee
The corresponding Riemann-Christoffel connection is 
\be
\Gam_{\lb\nu\mu} = - \frac{1}{3} \lh R_{\lb\mu\nu\kg} + R_{\nu\mu\lb\kg} \rh x^{\kg},
\label{6.3}
\ee
and indeed  $h_{\mu\nu}(0) = \Gam_{\lb\nu\mu}(0) = 0$. In a vacuum such a geometry is 
possible if $R_{\mu\nu\kg\lb} = W_{\mu\nu\kg\lb}$, i.e.\ if $R_{\mu\nu} = 0$. In the presence 
of matter it obviously requires a constant energy-momentum density: 
$\der_{\lb} T_{\mu\nu} = 0$. Clearly, a constant curvature automatically satisfies the 
Bianchi identity and the wave equatio, in agreement with equations (\ref{5.11})  and (\ref{5.12}).

A less trivial example is the asymptotic curvature of a spherically symmetric point-like test 
mass. In the linear approximation such mass may be modeled by a $\del$-function localizing 
the test mass on a world line $X^{\mu}(\tau)$; the corresponding energy-momentum tensor 
takes the form 
\be
T_{\mu\nu}(x) = m \int d\tau\, u_{\mu} u_{\nu}\, \del^4\left[ x - X(\tau) \right],
\label{6.4}
\ee
with $u^{\mu} = dX^{\mu}/d\tau$ the 4-velocity of the test mass. For a test mass at rest 
in the origin $u^{\mu} = (1,0,0,0)$ and the energy-momentum tensor reduces to a single 
component 
\be
T_{00}(x) = m\, \del^3(\bfx), \hs{2} \der_0 T_{00} = 0; \hs{3} T_{ij} = T_{i0} = 0.
\label{6.5}
\ee
Equivalently 
\be
\hT_{ij} = \del_{ij} \hT_{00} = \frac{1}{2}\, \del_{ij} T_{00}, \hs{2} \hT_{i0} = 0.
\label{6.5a}
\ee
Using the linearized Einstein equations it follows that 
\be
R_{ij} = \del_{ij} R_{00}, \hs{2} R_{00} = - 4 \pi G\, T_{00} = - 4 \pi G m\, \del^3(\bfx).
\label{6.6}
\ee
The full Riemann curvature tensor can be constructed by integrating equation 
(\ref{5.12}); using the retarded Green's function the solution of this equation for 
arbitrary $\hT_{\mu\nu}$ localized inside some finite volume $S$ becomes (\ref{5.6}):
\[
R_{\mu\nu\kg\lb} = \der_{\mu} \der_{\kg} \hh_{\nu\lb} - \der_{\mu} \der_{\lb} \hh_{\nu\kg} 
 - \der_{\nu} \der_{\kg} \hh_{\mu\lb} + \der_{\nu} \der_{\lb} \hh_{\mu\kg},
\]
where $\hh_{\mu\nu}$ is {\em defined} by
\be
\hh_{\mu\nu}(\bfx,t) = 2G\, \int_S d^3x'\, \frac{\hT_{\mu\nu}(\bfx',t_{ret})}{|\bfx - \bfx'|},
\label{6.7a}
\ee
the integrand being evaluated at the retarded time $t_{ret} = t - |\bfx - \bfx'|$. 
Notice that in view of equation (\ref{5.6}) $\hh_{\mu\nu}$ can be identified with a 
gauge-fixed expression for the metric fluctuation $h_{\mu\nu}$; however, as no gauge 
transformation applied to $h_{\mu\nu}$ will affect the curvature, it also will not affect
the result (\ref{6.7a}) for $\hh_{\mu\nu}$ for that matter. 

As the partial derivatives commute with $\Box^{-1}$ in Minkowski space, it follows that it
is possible to replace the terms in the expression for the Riemann curvature by the equivalent
\be
\der_{\mu}\der_{\kg} \hh_{\nu\lb}(\bfx,t) = 2G\, \int_S d^3x'\, 
 \frac{\left[\der'_{\mu} \der'_{\kg} \hT_{\nu\lb} \right](\bfx',t_{ret})}{|\bfx - \bfx'|}.
\label{6.7b}
\ee
Substitution of (\ref{6.5}), (\ref{6.5a}) into (\ref{6.7a}) or (\ref{6.7b}) now gives a direct 
expression for the curvature components, without requiring to fix a gauge for the tensor 
field $h_{\mu\nu}$.

The final result for the curvature field of the test mass $m$ now is the quadrupole field
\be
E_{ij} = - P_{ij} = \frac{3Gm}{r^5} \lh r_i r_i - \frac{1}{3}\, \del_{ij} r^2 \rh, \hs{2} B_{ij} = 0,
\label{6.8}
\ee
where $\bfer = \bfx - \bfx' \rightarrow \bfx$ if the test mass is located in the origin.
Equivalently,
\be
R_{ikjl} = \frac{3Gm}{r^5} \left[ \del_{ij} r_k r_l - \del_{il} r_k r_j - \del_{jk} r_i r_l + \del_{kl} r_i r_j 
 - \frac{2}{3}\, r^2 \lh \del_{ij} \del_{kl} - \del_{il} \del_{kj} \rh \right].
\label{6.9}
\ee
This is in full agreement with the asymptotic form for large $r$ of the Schwarzschild 
geometry.
\vs{2}

\np
\nit
{\bf \large 7.\ Gravitational radiation}
\vs{1}

\nit
We now turn to the more general asymptotic dynamical curvature solution of localized 
sources, the most important example of which is gravitational radiation\footnote{For a 
detailed and thorough introduction with references see \ct{maggiore:2008}.}. The general 
expression for the curvature in the weak-gravity limit is again (\ref{5.6}) for general 
time-dependent energy-momentum distributions $\hT_{\mu\nu}$ localized inside a finite 
volume $S$, which may for practical purposes be taken to be a sphere surrounding the 
source region; for non-relativistic sources it is convenient to take the center of mass of 
the source as the center of this sphere. 

As the linearized expression for $E_{ij} = R_{0i0j}$ reads
\be
E_{ij} = 2G \lh \der^2_0 h_{ij} - \der_0 \der_i h_{0j} - \der_0 \der_j h_{0i} + 
 \der_i \der_j h_{00} \rh,
\label{7.1e}
\ee
in the long-range limit $|\bfx - \bfx'| \simeq |\bfx| = r$ the leading terms in $1/r$ are
\be
\ba{lll}
E_{ij}(\bfx,t) & \simeq & \dsp{ \frac{2G}{r}\, \der_0^2 \int_S d^3x'\, \lh \hT_{ij} +\hr_i \hT_{0j}
 + \hr_j \hT_{0i} + \hr_i \hr_j \hT_{00} \rh (\bfx',t_{ret}) }\\
 & & \\
 & = & \dsp{ \frac{2G}{r}\, \der_0^2 \int_S d^3x'\, \left[ T_{ij} - \frac{1}{2} \lh \del_{ij} - \hr_i \hr_j \rh T_{kk}
  + \hr_i T_{0j} + \hr_j T_{0i} \rd }\\
 & & \\
 & & \dsp{ \hs{7} \ld +\, \frac{1}{2} \lh \del_{ij} + \hr_i \hr_j \rh T_{00} \right](\bfx',t_{ret}). }
\ea
\label{7.2e}
\ee
where $\hr_i = x_i/r$ are the components of the unit vector in the direction of $\bfx$.
Using a well-known identity for localized sources this can be written alternatively as
\be
\ba{lll}
E_{ij}(\bfx,t) & = & \dsp{ \frac{2G}{r} \left[ \frac{1}{2}\, \der_0^4 \int_S d^3x'\, x'_i x'_j T_{00}(\bfx',t_{ret}) - 
 \frac{1}{4} \lh \del_{ij} - \hr_i \hr_j \rh \der_0^4 \int_S d^3x'\, \bfx^{\prime\,2} T_{00}(\bfx',t_{ret}) \rd }\\
 & & \\
 & & \dsp{ \hs{2} +\, \hr_i\, \der^3_0 \int_S d^3x'\, x'_j\, T_{00}(\bfx',t_{ret}) + 
  \hr_j\, \der^3_0 \int_S d^3x'\, x'_i\, T_{00}(\bfx',t_{ret}) }\\
 & & \\
 & & \dsp{ \hs{2} \ld +\, \frac{1}{2} \lh \del_{ij} + \hr_i \hr_j \rh \der_0^2 \int_S d^3x' T_{00}(\bfx',t_{ret}) \right]. }\\
\ea
\label{7.3e}
\ee
Now in the non-relativistic limit all terms except the first two can be discarded, as they 
represent the time changes of the mass dipole and the mass monopole moment, which 
vanish in the CM frame. Thus we are left with 
\be
E_{ij}(\bfx,t) = \frac{G}{r}\, \der_0^4 \int d^3x' \left[ \lh x'_i x'_j - \frac{1}{3}\, \bfx^{\prime\,2} \del_{ij} \rh 
 - \frac{1}{2} \lh \hr_i \hr_j - \frac{1}{3}\, \del_{ij} \rh \bfx^{\prime\,2} \right] T_{00}(\bfx',t_{ret}).
\label{7.4e}
\ee
Note that the integral in the second term is equivalent to
\[
\der_0^4 \int_S d^3x'\, \bfx^{\prime\,2} T_{00} = 2\, \der_0^2 \int_S d^3x'\, T_{kk},
\]
which for non-relativistic sources is generally very small compared to the first term representing 
the mass quadrupole. Therefore in most practical situations the spatial trace-term can be
ignored. 

Although the expression (\ref{7.4e}) is manifestly traceless, as a result of the truncations made 
it is not manifestly transverse. This can be repared by considering the transverse projection 
of $E_{ij}$, which leads to the result:
\be
R_{0i0j}(\bfx,t) \simeq E_{ij}^{TT} = \lh \del_{ik} - \hr_i \hr_k \rh \lh \del_{jl} - \hr_j \hr_l \rh \lh E_{kl} + 
 \frac{1}{2}\, \del_{kl}\, \hr \cdot E \cdot \hr \rh. 
\label{7.5e}
\ee
Note that the $TT$-label here is not the result of gauge-fixing; it is simply a manifestly 
transverse and traceless set of components of the linearized Riemann --or equivalently 
linearized Weyl-- curvature tensor. 

This expression (\ref{7.5e}) is all that is needed to interpret the data of gravitational wave 
detectors; in general these monitor the space-time intervals between several test masses. 
If the space-time interval between two test masses moving on geodesics is $n^{\mu}$, the 
change in the interval due to space-time curvature is given by the geodesic deviation 
equation\footnote{For a derivation and discussion with references see \ct{vholten:2016}.}
\be
D_{\tau}^2 n^{\mu} = R_{\kg\nu\lb}^{\;\;\;\;\;\mu} u^{\kg} u^{\lb} n^{\nu}.
\label{7.6e}
\ee
where $D_{\tau}$ is a covariant proper-time derivative along the geodesic world line of one 
of the masses, $u^{\mu}$ is its four-velocity and $n^{\mu}$ measures the space-time interval 
of a second mass with respect to the first one. In the frame in which the first test mass is 
originally at rest in a local Minkowski frame: $u^{\mu} = (1,0,0,0)$, the equations simplify to 
\be 
\frac{d^2n^i}{d\tau^2} = E^{TT}_{ij} n^j, \hs{2} \frac{d^2n^0}{d\tau^2} = 0.
\label{7.7e}
\ee
Thus the relative motion of the test masses, determined by the curvature components 
$E_{ij}$, give direct access to the variations in the mass quadrupole moment of the source.
\vs{2}

\nit
{\bf \large 8.\ Beyond leading order}
\vs{1}

\nit
In the previous sections we derived expressions for the curvature induced by gravitational
fields at leading order. However, the expressions obtained there can be used as the 
starting point for an improved treatment of gravitational wave propagation including terms 
beyond leading order. The general procedure consists of expanding the Weyl tensor into 
the first-order contribution (\ref{5.6}) plus the next-order term: 
\be
W_{\mu\nu\kg\lb} = W^{(1)}_{\mu\nu\kg\lb} + W^{(2)}_{\mu\nu\kg\lb}, 
\label{8.1}
\ee
with $W^{(1)}_{\mu\nu\kg\lb}$ given by the linear approximation:
\[
W^{(1)}_{\mu\nu\kg\lb} = R^{(1)}_{\mu\nu\kg\lb} 
 = \der_{\mu} \der_{\kg} h^{(1)}_{\nu\lb} - \der_{\mu} \der_{\lb} h^{(1)}_{\nu\kg} 
 - \der_{\nu} \der_{\kg} h^{(1)}_{\mu\lb} + \der_{\nu} \der_{\lb} h^{(1)}_{\mu\kg},
\]
at the same time expanding the metric and the connection according to (\ref{5.1}):
\[
g_{\mu\nu} = \eta_{\mu\nu} + 2 h^{(1)}_{\mu\nu}, \hs{2} 
\Gam^{(1)}_{\lb\nu\mu} = \Gam_{\lb\nu}^{(1)\,\rg}\, \eta_{\rg\mu}
= \der_{\lb} h^{(1)}_{\nu\mu} + \der_{\nu} h^{(1)}_{\lb\mu} - \der_{\mu} h^{(1)}_{\lb\nu},
\]
where $h^{(1)}_{\mu\nu} = \hh_{\mu\nu}$ as given by eq.\ (\ref{6.7a}).

Armed with these results one now expands eq.\ (\ref{3.3}) as 
\be
\ba{l}
\Box W^{(2)}_{\mu\nu\kg\lb} = \nb^2_{(1)} W^{(1)}_{\mu\nu\kg\lb} + \eta^{\rg\eta} \eta^{\sg\tau}
 \lh 2 W^{(1)}_{\mu\sg\kg\rg} W^{(1)}_{\lb\eta\nu\tau} - W^{(1)}_{\mu\sg\lb\rg} W^{(1)}_{\kg\eta\nu\tau}
 + W^{(1)}_{\mu\nu\rg\sg} W^{(1)}_{\kg\lb\eta \tau}\rh \\
 \\
\hs{5} +\, \mbox{({\em source terms})},
\ea
\label{8.2}
\ee
where obviously the source terms are absent in the source-free far region. 
The first term on the right-hand side becomes to this approximation
\be
\ba{l}
\nb^2_{(1)} W^{(1)}_{\mu\nu\kg\lb} = \Box W^{(1)}_{\mu\nu\kg\lb} 
 + \eta^{\rg\eta} \eta^{\sg\tau} \left[ - 2 h^{(1)}_{\rg\sg} \der_{\eta} \der_{\tau} 
 W^{(1)}_{\mu\nu\kg\lb} - \Gam^{(1)}_{\tau\sg\rg}\, \der_{\eta} W^{(1)}_{\mu\nu\kg\lb} \rd \\
 \\
\hs{2} +\, 2 \lh \Gam^{(1)}_{\mu\sg\rg}\, \der_{\tau} W^{(1)}_{\eta\nu\kg\lb} + 
\Gam^{(1)}_{\nu\sg\rg}\, \der_{\tau} W^{(1)}_{\mu\eta\kg\lb} + \Gam^{(1)}_{\kg\sg\rg}\, \der_{\tau}
 W^{(1)}_{\mu\nu\eta\nu\lb} + \Gam^{(1)}_{\lb\sg\rg}\, \der_{\tau} W^{(1)}_{\mu\nu\kg\eta} \rh \\
 \\
\hs{2} \ld +\, (\der_{\tau} \Gam^{(1)}_{\mu\sg\rg}) W^{(1)}_{\eta\nu\kg\lb} + 
 (\der_{\tau} \Gam^{(1)}_{\nu\sg\rg}) W^{(1)}_{\mu\eta\kg\lb} +
 (\der_{\tau} \Gam^{(1)}_{\kg\sg\rg}) W^{(1)}_{\mu\nu\eta\lb} + 
 (\der_{\tau} \Gam^{(1)}_{\lb\sg\rg}) W^{(1)}_{\mu\nu\kg\eta} \right]. 
\ea
\label{8.3}
\ee
Of course the first term on the right-hand side here vanishes outside the source region, as by 
construction
\[
\Box W^{(1)}_{\mu\nu\kg\lb} = 0
\]
there. Clearly this perturbative scheme can be extended to still higher orders.
\vs{2}

\nit
{\bf \large 9.\ Discussion} 
\vs{1}

\nit
General Relativity is presently the best theory of gravity available, in agreement with 
almost all observations and experiments \ct{will:2014,ligo-virgo:2021}, with the possible 
exception of those phenomena which are usually interpreted as evidence for dark matter. 
Although all available tests are in the classical regime and we still do not have a complete  workable quantum theory of gravity, the observation of gravitational waves shows without 
doubt that in GR space-time is a genuine dynamical system. These waves in the 
space-time geometry are observed to propagate at the speed of light and have two 
transverse polarization modes. In view of the geometry of space-time being encoded 
in an observer-independent way by the curvature, it was the aim of this investigation 
to recast GR in such a way as to get a direct description of curvature dynamics. 

As the Ricci tensor and Riemann scalar are directly given in terms of the local 
energy-momentum distribution, the actual problem to be solved was to derive a 
wave equation for the purely gravitational degrees of freedom contained in the 
Weyl tensor. This has been achieved in equation (\ref{3.3}). It was also shown that 
in vacuum the propagating degrees of freedom of the Weyl tensor correspond to 
two independent transverse polarization modes, as was to be expected, and that in 
the non-relativistic linear approximation the curvature waves are sourced by the 
quadrupole modes of the energy-momentum distribution. 

Corrections to these results are of two kinds. First, the self-interaction of the gravitational 
field corrects the propagation of gravitational waves even on a flat background. As 
discussed in sect.\  8, the linearized curvature creates a non-flat background affecting 
the propagation of the non-linear second-order contributions to the gravitational waves. 
This will change the dispersion relation for the waves beyond first order.   

Another type of correction occurs if the gravitational fluctuations propagate in a 
non-vacuum environment. In the context of equation (\ref{3.3}) this means not 
only that there are non-vanishing source terms on the right-hand side, but also
that the 4-D laplacean $\nb^2$ is modified. This may manifest itself in additional
propagating degrees of freedom as $\nb^iE_{ij}$ and $\nb^i B_i^{\;j}$ receive 
contributions from the right-hand side of eq.\ (\ref{2.17}). These corrections 
require additional  study; they could show up e.g.\ in gravitational memory effects 
in non-flat background geometries \ct{bieri:2014, sarkkinen:2022}.
\vs{2}

\nit
{\bf \large Appendix A}
\vs{1}

\nit
This appendix addresses action principles for the Riemann en Weyl tensors. 
First observe that eq.\ (\ref{3.2}) can be rewritten using the Einstein tensor in 
the form
\be
\ba{l}
\nb^2 R_{\mu\nu\kg\lb} - 2 R_{\mu\sg\kg}^{\;\;\;\;\;\rg} R_{\lb\rg\nu}^{\;\;\;\;\;\sg} + 
2 R_{\mu\sg\lb}^{\;\;\;\;\;\rg} R_{\kg\rg\nu}^{\;\;\;\;\;\sg} + R_{\mu\nu\rg}^{\;\;\;\;\;\sg} 
R_{\kg\lb\sg}^{\;\;\;\;\;\rg} \\
 \\
\dsp{ \hs{1} +\, \frac{1}{2} \lh G_{\lb}^{\;\,\rg}  R_{\mu\nu\kg\rg} - G_{\kg}^{\;\rg} R_{\mu\nu\lb\rg}
 + G_{\nu}^{\;\rg} R_{\kg\lb\mu\rg} - R_{\mu}^{\;\rg} R_{\kg\lb\nu\rg} \rh - 
 2\, R\, R^{\mu\nu\kg\lb} R_{\mu\nu\kg\lb} }\\
 \\
\dsp{ \hs{1} =\, \frac{1}{2} \lh \left\{ \nb_{\mu}, \nb_{\kg} \right\} G_{\nu\lb} - 
 \left\{ \nb_{\mu}, \nb_{\lb} \right\} G_{\nu\kg} - \left\{ \nb_{\nu}, \nb_{\kg} \right\} G_{\mu\lb}
 + \left\{ \nb_{\nu}, \nb_{\lb} \right\} G_{\mu\kg} \rh }
\ea
\label{a.1}
\ee
For a fixed metric and connection, taking the tensor $R_{\mu\nu\kg\lb}$ as an independent 
set of variables, this equation defines an extremal point of the action functional 
\be
\ba{l}
S[R;g,H] = \dsp{ \int d^4x \sqrt{-g} \left[ - \frac{1}{2} \nb^{\sg} R^{\mu\nu\kg\lb} \nb_{\sg} 
 R_{\mu\nu\kg\lb} - \frac{1}{3} R_{\mu\nu}^{\;\;\;\kg\lb} R_{\kg\lb}^{\;\;\;\rg\sg} R_{\rg\sg}^{\;\;\;\mu\nu} 
 \rd }\\
 \\
\dsp{ \hs{2} \ld +\, \frac{4}{3}\, R_{\mu\nu}^{\;\;\;\kg\lb} R_{\kg\rg}^{\;\;\;\mu\sg} 
 R_{\;\,\sg\;\,\lb}^{\,\nu\;\,\rg} - 2\, G^{\rg}_{\;\,\lb} R^{\mu\nu\kg\lb} R_{\mu\nu\kg\rg} - 
 R\, R^{\mu\nu\kg\lb} R_{\mu\nu\kg\lb} - H_{\mu\nu\kg\lb} R^{\mu\nu\kg\lb} \right], }
\ea
\label{a.2}
\ee
where using the Einstein equations the source term can be replaced by
\be
H_{\mu\nu\kg\lb} = - 4\pi G \lh \left\{ \nb_{\mu}, \nb_{\kg} \right\} T_{\nu\lb} - 
 \left\{ \nb_{\mu}, \nb_{\lb} \right\} T_{\nu\kg} - \left\{ \nb_{\nu}, \nb_{\kg} \right\} T_{\mu\lb}
 + \left\{ \nb_{\nu}, \nb_{\lb} \right\} T_{\mu\kg} \rh.
\label{a.3}
\ee
For the case of vacuum solutions with $G_{\mu\nu} = R = H_{\mu\nu\kg\lb} = 0$ the 
action $S[R;g,h]$ straightforwardly reduces to the action $S[W;g]$ in eq.\ (\ref{3.4}) 
for the corresponding traceless part $W_{\mu\nu\kg\lb}$ of the tensor $R_{\mu\nu\kg\lb}$. 
Obviously, for the given metric and connection solutions of (\ref{a.1}) for the tensor 
$R_{\mu\nu\kg\lb}$ are given by equation (\ref{2.2}). However, the example of deriving 
curvature fluctuations directly in the linearized theory shows, that the alternative approach 
presented here is feasible for certain applications, which could e.g.\ include the related 
problem in non-vacuum space-times with additional matter fields.

The observation that this equation defines an extremum of the functional (\ref{a.2}) 
can be useful in such cases, for example to derive WKB-type of approximations. 
More generally, for given background $(G^{\mu}_{\;\,\nu}, R)$, as defined by the 
energy-momentum tensor, equations (\ref{a.1}) and (\ref{a.2}) relate fluctuations 
of the curved-space d'Alembertian $\nb^2$ to curvature fluctuations, thereby connecting 
two different approaches --spectral analysis and differential geometry-- to the description 
and analysis of gravitational fluctuations.

\vs{3}

\nit
{\bf Acknowledgement} \\
The author is indebted to Miika Sarkkinen of the University of
Helsinki for correspondence.

\end{document}